# A SOCP-based ACOPF for Operational Scheduling of Three-phase Unbalanced Distribution Systems and Coordination of PV Smart Inverters

Mingyue He, *Student Member, IEEE*, Zahra Soltani, *Student Member, IEEE*, Mohammad Ghaljehei, *Student Member, IEEE*, Masoud Esmaili, *Senior Member, IEEE*, Shanshan Ma, *Member, IEEE*, Mengxi Chen, *Student Member, IEEE*, Mojdeh Khorsand, *Member, IEEE*, Raja Ayyanar, *Member, IEEE*, Vijay Vittal, *Life Fellow, IEEE*

*Abstract* — The proliferation of distributed energy resources (DERs) imposes new challenges to distribution system operation, e.g., power quality issues. To overcome these challenges and enhance system operation, it is critical to effectively utilize all available resources and accurately characterize unbalanced distribution networks in operational tools. This paper proposes a convex second-order-cone programming (SOCP)-based AC optimal power flow (ACOPF) model for three-phase unbalanced distribution networks, including smart inverters and Volt-VAr controller (VVC) devices. Reactive power-voltage (Q-V) characteristics of smart inverters of solar photovoltaic (PV) units are also modeled. Moreover, the settings of Q-V characteristics of VVC are co-optimized within the proposed ACOPF, considering the allowable range of the IEEE 1547-2018 standard. Furthermore, dynamic analyses are conducted to verify the stability of optimal settings of VVC. The proposed models are tested on an actual 1747-node primary distribution feeder in Arizona. The results illustrate the effectiveness of the proposed ACOPF for unbalanced systems in providing a global optimal solution while capturing the non-linearity and non-convexity of ACOPF. By co-optimizing settings, system operation is improved due to the flexibility of adjusting reactive power output from PV units with VVC. The time-domain simulations show that the optimal settings result in a stable system.

## Nomenclature

*Indices and Sets*

| | |
|---|---|
| $d \in D$ | Index and set of demands. |
| $g \in G$ | Index and set of nodes connected to substations. |
| $h \in H$ | Index and set of all solar photovoltaic (PV) units. |
| $i\phi \in B$ | Index and set of nodes (bus $i$ and phase $\phi$). |
| $k \in K$ | Index and set of all PV units having Volt-VAr controller (VVC). |
| $l \in L$ | Index and set of lines. |
| $n \in N$ | Index and set of voltage settings of VVC. |
| $\phi \in \Phi$ | Index and set of phases. |
| $G(i\phi)$ | Sets of connections for substation at node $i\phi$. |
| $H(i\phi)$ | Sets of connections for PV unit at node $i\phi$. |
| $B(i\phi)$ | Sets of line connections for node $i\phi$. |
| $\mathcal{D}(h)$ | Sets of the load node connected to PV unit $h$. |

*Parameters*

| | |
|---|---|
| $\rho^G, \rho^{pv}$ | Energy prices from bulk system and PV units. |
| $v_n$ | Auxiliary parameter representing square of default voltage setting $n$ of VVC. |
| $v_{n,k}^{(t)}$ | Taylor series first-order expansion base point for auxiliary parameter representing square of voltage setting $n$ for PV unit $k$ with VVC at iteration $t$. |
| $\theta_{i\phi}^{(t)}$ | Taylor series first-order expansion base point for voltage angle at node $i\phi$ at iteration $t$. |
| $b_{ij}^{\phi\phi'}$ | Susceptance between node $i\phi$ and node $j\phi'$. |
| $g_{ij}^{\phi\phi'}$ | Conductance between node $i\phi$ and node $j\phi'$. |
| $u_{i\phi}^{(t)}$ | Taylor series first-order expansion base point for auxiliary variable $u_{i\phi}$ at node $i\phi$ at iteration $t$. |
| $P_d^D, Q_d^D$ | Active and reactive power demands at node $d$. |
| $P_{pv,h}^{max}$ | Maximum active power point of PV unit $h$. |
| $Q_{pv,k}^{max}$ | Maximum reactive power setting for PV unit $k$ with VVC. |
| $S_{pv,h}^{max}$ | Apparent power rating of PV unit $h$. |
| $V_{i\phi}^L, V_{i\phi}^U$ | Minimum and maximum of voltage magnitude limits at node $i\phi$. |

*Decision Variables*

| | |
|---|---|
| $\theta_{i\phi}$ | Voltage angle at node $i\phi$. |
| $\tilde{v}_{n,k}$ | Auxiliary variable representing square of voltage setting $n$ of Q-V curve of PV unit $k$ with VVC. |
| $c_{ij}^{\phi\phi'}, e_{ij}^{\phi\phi'}$ | Auxiliary variables for SOCP-based ACOPF. |
| $u_{i\phi}$ | Auxiliary variable for SOCP-based ACOPF. |
| $z_{n,k}$ | Binary variable for operating zone $n$ of VVC for PV unit $k$. |
| $P_g^G, Q_g^G$ | Active and reactive powers from upstream system in node $g$. |
| $P_l^L, Q_l^L$ | Active and reactive power flows of branch $l$. |
| $P_h^{PV}, Q_h^{PV}$ | Active and reactive powers of PV unit $h$. |
| $\tilde{Q}_{pv,k}^{max}$ | Maximum reactive power setting for PV unit $k$ with VVC. |

## I. Introduction

### A. Motivation and Background

The increasing penetration level of distributed energy resources (DERs) has made them an indispensable part of modern distribution network operation. Different IEEE Standards have been established to enhance power system operation and overcome challenges associated with integrating DERs in grid operation. For example, Standards Coordinating Committee 21

Authors are with the Department of Electrical, Computer and Energy Engineering, Arizona State University, Tempe, AZ 85281 USA (e-mail: mingyue1@asu.edu, zsoltani@asu.edu, mghaljeh@asu.edu, msdesmaili@gmail.com, shansh19@asu.edu, Mengxi.Chen@asu.edu, mojdeh.khorsand@asu.edu, rayyanar@asu.edu, Vijay.Vittal@asu.edu).

This research is funded by the Department of Energy (DOE) Advanced Research Projects Agency – Energy (ARPA-E) under OPEN 2018 program Award DE-AR0001001.

(SCC21) has developed IEEE Standard 1547-2018 for the interconnection and interoperability of DERs in distribution systems [1]. This standard provides a uniform criterion and sets the requirements relevant to the performance, operation testing, safety considerations, and maintenance of the interconnection of DERs such that they can be universally adopted. With the issued IEEE Standard 1547-2018, there is a need to update system operational tools to improve the representation of DERs in distribution operation.

Rooftop solar photovoltaic (PV) unit is one type of DER, which has been widely adopted in distribution systems. When operated in the Reactive power-voltage (Q-V) controller mode, Volt-VAr controllers (VVC) of smart inverters can adjust the power outputs of PV units and maintain the voltage within an acceptable range in [1]. For example, excessive PV generation in distribution systems can lead to overvoltage issues, which can be mitigated via two options: (i) reactive power support under Q-V mode and (ii) active power curtailment of PV units. Thus, PV units with smart inverters constitute a set of dispatchable resources whose optimal operation can enhance the performance of distribution networks.

As a result of the high R/X ratio of distribution feeders, the assumption of DC optimal power flow typically made in transmission systems is not valid. Moreover, unlike balanced transmission grids, the distribution networks are commonly unbalanced, and the mutual coupling between the three phases cannot be ignored. Therefore, a three-phase unbalanced AC optimal power flow (ACOPF) model is essential for distribution system operational scheduling. However, the three-phase unbalanced ACOPF is highly nonlinear and non-convex. Directly solving a nonlinear and non-convex model is not preferred because of the quality of the solution and the computational requirements.

Therefore, with the increasing penetration of DERs and the recommendations of IEEE standard 1547-2018, there is an urgent need to formulate a convex three-phase unbalanced ACOPF that incorporates the dispatching and performance of DERs in distribution grids.

*B. Literature Review*

Numerous studies have been conducted for the convexification of nonlinear and non-convex ACOPF. For the balanced system, two main convex relaxation techniques, which are semidefinite programming (SDP) [2]-[6] and second-order-cone programming (SOCP) [7]-[13], are explored to obtain the global optimal solution of ACOPF. Reference [2] is among the first papers that proposed SDP-based ACOPF and solved it using the interior point method (IPM) algorithm. Reference [3] implements three decomposition techniques to decrease the computational time of the SDP-based ACOPF approach. Stronger and tighter SDP relaxation is discussed in [4]-[6] to mitigate the non-exact issue of the SDP relaxation. The ACOPF is first convexified and reformed as a SOCP problem for radial networks in [7]. Reference [8] uses hierarchies of linear programming with SOCP to alleviate the computational burden. Reference [9] utilizes a SOCP-based ACOPF to obtain optimal online control of devices. A mixed-integer SOCP problem is presented for the reactive optimal power flow (OPF) to determine the status of shunt elements and tap ratio of transformers in [10]. Another mixed-integer SOCP model is proposed by [11] to alleviate the unbalance issue from the demand-side with DERs. In [12], SOCP-based ACOPF is used in a security-constrained OPF problem for the worst contingencies of the system. The authors in [13] use a SOCP-based model to improve the grid operation with DERs for balanced systems. However, the primary and low-voltage level distribution girds are generally unbalanced due to unbalanced loads, DERs, and line segments. The assumptions of ACOPF in [2]-[13] may not be accurate enough to address the condition of the system and ensure the secure and economic operation in distribution grid.

Recently, some research efforts have been conducted to formulate the convexified three-phase unbalanced ACOPF model. The SDP-based three-phase unbalanced ACOPF model in [14] is proposed and solved by alternating direction method of multipliers (ADMM). Reference [15] proposes a chordal relaxation-based SDP model for ACOPF in unbalanced systems considering DERs and voltage regulation transformers (VRTs) and provides a tighter convex model for VRTs to mitigate solution inexactness. Reference [16] convexifies nonlinear three-phase unbalanced ACOPF through a moment relaxation-based SDP model with a two-stage hierarchical algorithm to obtain the exact feasible solution. The SDP-based ACOPF for an unbalanced system in [17] accounts for the mixture of wye and delta-connection loads, DERs, and step voltage regulators. All references [14]-[17] utilize the SDP technique to convexify the ACOPF. The SOCP-based model is another potential option for convexifying ACOPF for three-phase unbalanced systems, but rarely discussed in literature. To fill this gap, this paper proposes a convex SOCP-based ACOPF model for unbalanced distribution grids, and the performance of the proposed method is evaluated an actual distribution primary feeder.

The modeling of PV units in the distribution grid operation has been addressed in the literature. In [18]-[19], the active and reactive power and power factor limits of PV units are considered in OPF. Reference [20] proposes three different control strategies with active and reactive power limits for PV units to improve the voltage profile in distribution networks. However, the PV models in [18]-[20] may not satisfy IEEE standard 1547-2018 recommendations. The DERs standard performance in IEEE standard 1547-2018 is studied in [21]. The author formulates a decentralized approach to account for the standard characteristic in IEEE standard 1547-2018 for balanced systems. Reference [22] proposed a two-level Volt-VAr control scheme of PV units in which a 15-min dispatch and real-time adjustment are considered for PV units to enhance system operation. The unbalanced three-phase distribution grid characteristics are not well captured in [21]-[22]. To improve the modeling of DERs in system operational tools in unbalanced distribution grids, this paper incorporates the IEEE standard 1547-2018 characteristic of PV units with a mixed-integer SOCP (MISOCP)-based ACOPF. Two different models of settings of VVC are considered: fixed default settings and optimal settings.

*C. Contributions and Organization of the Paper*

According to the gaps mentioned above, the main contributions of this paper can be summarized as follows:

- A convex SOCP-based ACOPF which accounts for the unique characteristics of distribution networks, i.e., single-phase and three-phase lines, unbalanced network, and mutual impedances between phases, is proposed for three-phase unbalanced distribution systems. A two-stage algorithm is developed to solve the proposed model. The convex SOCP-based ACOPF is tested on a real 1747-node primary distribution feeder in Arizona with significant roof-top PV penetration. The simulation results indicate that the proposed model has the ability to capture the three-phase unbalanced characteristics of distribution grids and obtain a global optimal solution that is exact for the distribution system.

- The proposed SOCP-based ACOPF is extended and converted into a MISOCP-based ACOPF model to account for the Q-V characteristics of PV units equipped with VVCs based on the IEEE standard 1547-2018. Two different types of models for VVC are studied: (i) default settings and (ii) optimal settings. The simulation results show that the proposed MISOCP-based ACOPF model can capture the Q-V characteristic of VVC and the characteristics of the unbalanced network.
- This work also shows the advantages of optimal settings of VVC in comparison with default settings. Case studies show that co-optimizing the settings of VVC within the allowable range of IEEE standard 1547-2018 in the ACOPF model enables more flexibility to adjust the reactive power output of PV units with VVC, which can mitigate voltage issues, improve system operation, and reduce operational cost. Dynamic analyses are conducted to verify the control stability of PV units under optimal settings of VVCs and load changes. The time-domain simulations show that the decision of optimal settings is valid and does not result in any stability issues in the system.

Section II explains the formulations and the algorithm of the proposed convex SOCP-based ACOPF. The Q-V characteristic of PV units and corresponding formulations in MISOCP-based ACOPF are introduced in Section III. Numerical results of the proposed model and dynamic simulation are presented and discussed in Section IV. Finally, Section V concludes the paper.

## II. CONVEX ACOPF FOR THREE-PHASE UNBALANCED DISTRIBUTION SYSTEMS

### A. Convex SOCP-based ACOPF for Unbalanced Systems

The complex power flow on the line $\boldsymbol{S}_{ij\phi}^L$ from bus $i$ to bus $j$ at phase $\phi$ can be calculated through the voltage $\boldsymbol{V}_{i\phi}$ multiplied by the conjugate of the line current (phasors are in bold font). The formulation of power flow of a three-phase line is shown in (1).

$$\boldsymbol{S}_{ij\phi}^L = \boldsymbol{V}_{i\phi}\left(\sum_{\phi'\in\Phi}(\boldsymbol{V}_{i\phi'}-\boldsymbol{V}_{j\phi'})\boldsymbol{Y}_{ij}^{\phi\phi'}\right)^* \quad (1)$$

where $\boldsymbol{V}_{i\phi}$ is the voltage at bus $i$ on phase $\phi$; $\boldsymbol{Y}_{ij}^{\phi\phi'}$ is the admittance of the path from bus $i$ phase $\phi$ to bus $j$ phase $\phi'$; note that $\boldsymbol{Y}_{ij}^{\phi\phi}$ indicates self-impedance of phase $\phi$ and $\boldsymbol{Y}_{ij}^{\phi\phi'}$ ($\phi \ne \phi'$) indicates mutual impedances between phases $\phi$ and $\phi'$ with phasor $\boldsymbol{V}_{i\phi} = V_{i\phi}\angle\theta_{i\phi}$ and $\boldsymbol{Y}_{ij}^{\phi\phi'} = g_{ij}^{\phi\phi'} + jb_{ij}^{\phi\phi'}$. $V_{i\phi}$ and $\theta_{i\phi}$ are the voltage magnitude and the voltage angle at bus $i$ phase $\phi$ respectively. By substituting $\boldsymbol{V}_{i\phi}$ and $\boldsymbol{Y}_{ij}^{\phi\phi'}$ in (1), $\boldsymbol{S}_{ij}^L$ can be written as (2).

$$\boldsymbol{S}_{ij\phi}^L = V_{i\phi}V_{i\phi}(g_{ij}^{\phi\phi}-jb_{ij}^{\phi\phi}) + \sum_{\phi'\in\Phi\backslash\phi}V_{i\phi}V_{i\phi'}\left(\cos\theta_{ii}^{\phi\phi'} + j\sin\theta_{ii}^{\phi\phi'}\right)\left(g_{ij}^{\phi\phi'}-jb_{ij}^{\phi\phi'}\right) - \sum_{\phi'\in\Phi}\left(\cos\theta_{ij}^{\phi\phi'} + j\sin\theta_{ij}^{\phi\phi'}\right)V_{i\phi'}V_{j\phi'}\left(g_{ij}^{\phi\phi'}-jb_{ij}^{\phi\phi'}\right) \quad (2)$$

Three auxiliary variables (3)-(5) are introduced to reformulate the nonlinear three-phase unbalanced formulation (2) and eventually convexify it in (6)-(8) and (10).

$$c_{ij}^{\phi\phi'} = V_{i\phi}V_{j\phi'}\cos\theta_{ij}^{\phi\phi'} \quad (3)$$

$$e_{ij}^{\phi\phi'} = V_{i\phi}V_{j\phi'}\sin\theta_{ij}^{\phi\phi'} \quad (4)$$

$$u_{i\phi} = (V_{i\phi})^2 \quad (5)$$

By substituting three auxiliary variables into (2), the active and reactive power flows on the line from bus $i$ to bus $j$ on phase $\phi$ can be reformulated as a convex form (6)-(7). The relationship of the auxiliary variables $c_{ij}^{\phi\phi'}$ and $s_{ij}^{\phi\phi'}$ are presented via (8)-(9).

$$P_{ij\phi}^L = g_{ij}^{\phi\phi}u_{i\phi} + \sum_{\phi'\in\Phi\backslash\phi}\left(g_{ij}^{\phi\phi'}c_{ii}^{\phi\phi'} + b_{ij}^{\phi\phi'}e_{ii}^{\phi\phi'}\right) - \sum_{\phi'\in\Phi}\left(g_{ij}^{\phi\phi'}c_{ij}^{\phi\phi'} + b_{ij}^{\phi\phi'}e_{ij}^{\phi\phi'}\right), \forall ij\phi \in L \quad (6)$$

$$Q_{ij\phi}^L = -b_{ij}^{\phi\phi}u_{i\phi} + \sum_{\phi'\in\Phi\backslash\phi}\left(g_{ij}^{\phi\phi'}e_{ii}^{\phi\phi'} - b_{ij}^{\phi\phi'}c_{ii}^{\phi\phi'}\right) - \sum_{\phi'\in\Phi}\left(g_{ij}^{\phi\phi'}e_{ij}^{\phi\phi'} - b_{ij}^{\phi\phi'}c_{ij}^{\phi\phi'}\right), \forall ij\phi \in L \quad (7)$$

$$c_{ij}^{\phi\phi'} = c_{ji}^{\phi'\phi}, \ e_{ij}^{\phi\phi'} = -e_{ji}^{\phi'\phi} \quad (8)$$

$$\left(c_{ij}^{\phi\phi'}\right)^2 + \left(e_{ij}^{\phi\phi'}\right)^2 = u_{i\phi}u_{j\phi'} \quad (9)$$

The constraint (9) shows the exact relationship between auxiliary variables; however, it introduces non-convexity in the model. Therefore, the non-convex constraint (9) is relaxed as a convex SOC constraint in (10).

$$\left(c_{ij}^{\phi\phi'}\right)^2 + \left(e_{ij}^{\phi\phi'}\right)^2 \le u_{i\phi}u_{j\phi'} \quad (10)$$

A general three-phase unbalanced ACOPF can be formulated as a convex SOCP-based optimization problem as shown in (11)-(18) with the core constraints (6)-(8), (10). In this general model, the PV units without VVC are non-controllable. As a result, these PV units inject maximum available active power with zero reactive power into the distribution grid.

$$\min \sum_{\forall g\in G}\rho^G P_g^G + \sum_{\forall h\in H}\rho^{pv}\left(P_h^{pv} - P_{i\phi, \ if \ i\phi \ \in \mathcal{D}(h)}^D\right) \quad (11)$$

$$\sum_{\forall g\in G(i\phi)}P_g^G + \sum_{\forall h\in H(i\phi)}P_h^{pv} = \sum_{\forall i\phi\in D}P_{i\phi}^D + \sum_{\forall j\in B(i\phi)}P_{ij\phi}^L, \forall i\phi \in B \quad (12)$$

$$\sum_{\forall g\in G(i\phi)}Q_g^G + \sum_{\forall h\in H(i\phi)}Q_h^{pv} = \sum_{\forall i\phi\in D}Q_{i\phi}^D + \sum_{\forall j\in B(i\phi)}Q_{ij\phi}^L, \forall i\phi \in B \quad (13)$$

$$P_h^{pv} = P_{pv,h}^{max}, \forall h \in H \quad (14)$$

$$Q_h^{pv} = 0, \forall h \in H \quad (15)$$

$$(V_{i\phi}^L)^2 \le u_{i\phi} \le (V_{i\phi}^U)^2, \forall i\phi \in B\backslash G \quad (16)$$

$$u_g = (V_g)^2, \forall g \in G \quad (17)$$

$$\theta_g = \theta_g^{sub}, \forall g \in G \quad (18)$$

The objective function (11) minimizes the total system operating cost including the cost of energy purchased from the upstream wholesale market and the cost of solar generation surplus paid to PV owners. The active and reactive power balances at each node are constrained by (12)-(13). The VVC enables changing the active and reactive power output of PV units and provide voltage regulation service. For the PV units without VVC, the PV units' active and reactive outputs are modeled as (14)-(15). Voltage limits of nodes are given by (16). Constraints (17)-(18) fix voltage magnitude and angle for nodes connected to the substation.

However, the reformed formulations (6)-(8), (10) provide no guarantee of a correct and exact solution. One reason is that the mutual impedance of the three-phase line creates the virtual loops shown in Fig. 1, which makes the network non-radial and complicates the determination of an exact solution using the SOCP-based formulations (6)-(8) and (10).

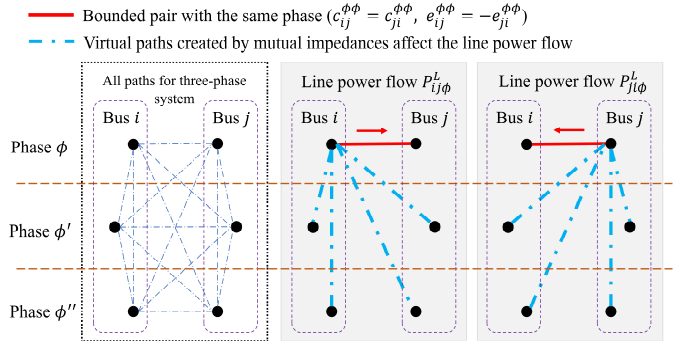

Fig. 1. Paths and line power flows for the three-phase unbalanced network.

Another reason is that the active and reactive power flow constraints are not completely bounded by the auxiliary variable constraint (8). For example, unlike SOCP-based ACOPF constraints in the balanced system, the reformed formulations (6)-(8) and (10) cannot ensure that the active power line loss is nonnegative in the unbalanced distribution system. In the nonlinear and non-convex line flow constraint (2), based on trigonometry, $\cos\theta_{ij}^{\phi\phi'}$ and $\sin\theta_{ij}^{\phi\phi'}$ between different phase pairs have a strong correlation. However, the constraint (8) only ensures the correlation of auxiliary variables that are related to the phase pairs with the same phase in line flow formulation (6). For example, the line power flow $P_{ij\phi}^L$ and $P_{ji\phi}^L$ are affected by both actual and virtual paths due to the mutual impedance shown in Fig. 1. Only the red path for the similar phase is bounded to ensure the correlation in the line flow constraints (e.g., $c_{ij}^{\phi\phi} = c_{ji}^{\phi\phi}$, $e_{ij}^{\phi\phi} = -e_{ji}^{\phi\phi}$). In contrast, the terms with auxiliary variables related to phase pairs with different phases (e.g., phase $\phi$ and phase $\phi'$ for $c_{ij}^{\phi\phi'}$ in $P_{ij\phi}^L$ and $c_{ji}^{\phi\phi'}$ in $P_{ji\phi}^L$) have no limit and mathematical correlation for line power flow constraints. Those unbounded auxiliary variable pairs make it rather hard to obtain an exact solution. To overcome the challenges mentioned above, new bounding constraints are proposed for the three-phase unbalanced SOCP-based ACOPF and solved by a two-stage algorithm.

*B. New Bounding Constraints and Two-Stage Algorithm for Solving Three-Phase Unbalanced SOCP-based ACOPF*

This paper proposes new bounding constraints for auxiliary variables $c_{ij}^{\phi\phi'}$ and $e_{ij}^{\phi\phi'}$ to narrow down the feasible region, address the problems of virtual paths created by mutual impedances, and bound the terms associated with different phases via Taylor series approximation. The formulations (19)-(20) are linearized expressions of auxiliary variables $c_{ij}^{\phi\phi'}$ and $e_{ij}^{\phi\phi'}$ in (3)-(4) by applying the first-order Taylor series approximation.

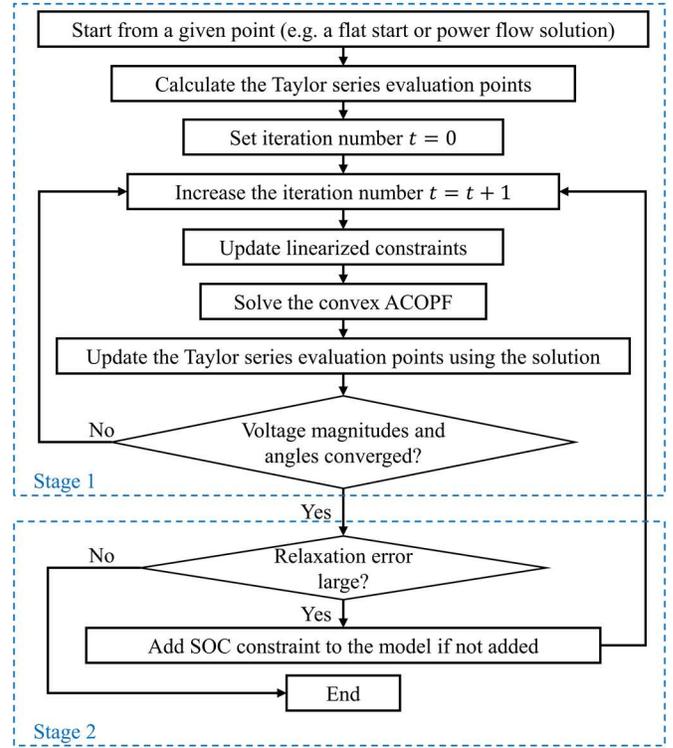

Fig. 2. The proposed two-stage algorithm to solve the proposed SOCP-based ACOPF for three-phase unbalanced systems.

$$c_{ij}^{\phi\phi'} = \sqrt{u_{i\phi}^{(t)} u_{j\phi'}^{(t)}} \left(\theta_{i\phi}^{(t)} - \theta_{j\phi'}^{(t)}\right) \sin\left(\theta_{i\phi}^{(t)} - \theta_{j\phi'}^{(t)}\right) +$$
$$\frac{\sqrt{u_{j\phi'}^{(t)}}}{2\sqrt{u_{i\phi}^{(t)}}} \cos\left(\theta_{i\phi}^{(t)} - \theta_{j\phi'}^{(t)}\right) u_{i\phi} + \frac{\sqrt{u_{i\phi}^{(t)}}}{2\sqrt{u_{j\phi'}^{(t)}}} \cos\left(\theta_{i\phi}^{(t)} - \theta_{j\phi'}^{(t)}\right) u_{j\phi'} -$$
$$\sqrt{u_{i\phi}^{(t)} u_{j\phi'}^{(t)}} \sin\left(\theta_{i\phi}^{(t)} - \theta_{j\phi'}^{(t)}\right) (\theta_{i\phi} - \theta_{j\phi'}) \quad (19)$$

$$e_{ij}^{\phi\phi'} = -\sqrt{u_{i\phi}^{(t)} u_{j\phi'}^{(t)}} \left(\theta_{i\phi}^{(t)} - \theta_{j\phi'}^{(t)}\right) \cos\left(\theta_{i\phi}^{(t)} - \theta_{j\phi'}^{(t)}\right) +$$
$$\frac{\sqrt{u_{j\phi'}^{(t)}}}{2\sqrt{u_{i\phi}^{(t)}}} \sin\left(\theta_{i\phi}^{(t)} - \theta_{j\phi'}^{(t)}\right) u_{i\phi} + \frac{\sqrt{u_{i\phi}^{(t)}}}{2\sqrt{u_{j\phi'}^{(t)}}} \sin\left(\theta_{i\phi}^{(t)} - \theta_{j\phi'}^{(t)}\right) u_{j\phi'} +$$
$$\sqrt{u_{i\phi}^{(t)} u_{j\phi'}^{(t)}} \cos\left(\theta_{i\phi}^{(t)} - \theta_{j\phi'}^{(t)}\right) (\theta_{i\phi} - \theta_{j\phi'}) \quad (20)$$

Since Taylor series first-order expansion base points (e.g., $u_{i\phi}^{(t)}$ and $\theta_{i\phi}^{(t)}$) are introduced in (19)-(20), those base points need to be iteratively updated in the SOCP-based ACOPF model to obtain an exact and feasible solution. In this regard, a two-stage algorithm is developed to solve the proposed three-phase unbalanced SOCP-based ACOPF model with new bounding constraints (19)-(20).

The overall two-stage algorithm is presented in Fig. 2. The algorithm updates the Taylor series base points and adds SOC

constraints if needed. Stage 1 starts from an initial point, which can be a flat start for all nodes with voltage magnitudes of the substation and voltage angles 0°, -120°, +120° for phases $a$, $b$ and $c$, respectively. Moreover, a power flow solution of voltage magnitudes and angles from a similar system condition can be used as an initial point. With the selected initial point, the initial Taylor series base points (i.e., $u_{i\phi}^{(t)}$ and $\theta_{i\phi}^{(t)}$) can be calculated for the linearized constraints (19)-(20). Then, Stage 1 goes into the loop to make the voltage magnitudes and angles converged. In the first iteration of the loop, the initial Taylor series base points are used to update linearized constraints (19)-(20) and the convex SOCP-based ACOPF (6)-(8), (11)-(20) is solved. After obtaining the solution of the convex ACOPF model (6)-(8), (11)-(20), new Taylor series base points are calculated based on the obtained solution to prepare for the update in the next iteration. If the updated voltage magnitudes and angles are close enough to the previous values between iterations, the update of the Taylor series base points has converged. In that case, the algorithm can proceed to Stage 2. Otherwise, Stage 1 continues to update Taylor series base points in the loop until they meet the tolerance criterion. In Stage 2, the absolute SOC relaxation errors $u_{i\phi}u_{j\phi'} - \left(c_{ij}^{\phi\phi'}\right)^2 - \left(e_{ij}^{\phi\phi'}\right)^2$ are checked. If the absolute relaxation errors are minor, the algorithm stops and outputs the obtained solution. Otherwise, the SOC constraint (10) is added, and the algorithm goes back to Stage 1.

### III. Q-V Characteristic of PV Units Modeled in ACOPF

#### A. Q-V Characteristic of Photovoltaic Units with VVC

In the previous section, the convex SOCP-based ACOPF is proposed with the non-dispatchable PV units that are not equipped with VVC. This section extends the proposed ACOPF model to account for Q-V characteristics of PV units with VVC to enable effective operational scheduling of these resources. To this end, the SOCP-based ACOPF model is modified to a MISOCP-based ACOPF. Moreover, two different models are developed for the default and optimal settings of VVC shown in Section III.B and Section III.C, respectively.

IEEE standard 1547-2018 indicates that the DERs with VVC should be capable of injecting and absorbing reactive power and participating in voltage regulation [1]. Furthermore, this standard provides four general modes for the reactive power control functions of DERs [1]. This paper considers the Q-V mode presented in Fig. 3, which is a mandatory requirement for both Category A and Category B 1547-compliant inverters.

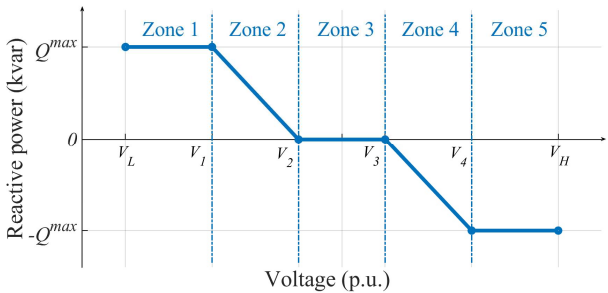

Fig. 3. Voltage-reactive power mode: Q-V characteristic of PV units with five different operating zones.

$V_L$ and $V_H$ are the lower and upper limits for DER continuous operation in Fig. 3. $Q^{max}$ is the setting for the maximum reactive power output of DER. As the auxiliary variable $u_{i\phi}$ is defined to represent squared voltage magnitude in the proposed SOCP-based ACOPF, the voltage settings for the separation of different operating zones $V_1$, $V_2$, $V_3$, $V_4$, $V_L$ and $V_H$ are squared as $v_1$, $v_2$, $v_3$, $v_4$, $v_L$ and $v_H$, respectively. The mathematical expression of the Q-V characteristic can be expressed as (21).

$$Q_k^{pv} = \begin{cases} Q_{pv,k}^{max} & \text{if } v_L \leq u_k < v_1 \\ \frac{Q_{pv,k}^{max}}{v_2 - v_1}(v_2 - u_k) & \text{if } v_1 \leq u_k < v_2 \\ 0 & \text{if } v_2 \leq u_k < v_3, \forall k \in K \quad (21) \\ \frac{Q_{pv,k}^{max}}{v_3 - v_4}(u_k - v_3) & \text{if } v_3 \leq u_k < v_4 \\ -Q_{pv,k}^{max} & \text{if } v_4 \leq u_k < v_H \end{cases}$$

The output of PV units, which are equipped with VVC, can be controlled and adjusted. The active and reactive power output limits of PV units with VVC are modeled as (22)-(23).

$$\left(P_k^{pv}\right)^2 + \left(Q_k^{pv}\right)^2 \leq \left(S_{pv,k}^{max}\right)^2, \forall k \in K \quad (22)$$

$$P_k^{pv} \leq P_{pv,k}^{max}, \forall k \in K \quad (23)$$

#### B. Formulations of Q-V Characteristic of Photovoltaic Units in ACOPF with Default Settings

The expression of the Q-V characteristic in (21) is piecewise linear. In this section, first, the expression (21) is converted into five different operating zones using a linear representation. Binary variables, i.e., $z_{1,k}$, $z_{2,k}$, $z_{3,k}$, $z_{4,k}$, and $z_{5,k}$ are introduced for Zones 1-5, respectively. The Big-M method is used to formulate if-then conditional statements for the five operating zones. The square voltage settings $v_1$, $v_2$, $v_3$, and $v_4$ of VVC are fixed default values based on IEEE standard 1547-2018, which are $0.94^2$, $0.98^2$, $1.02^2$, and $1.06^2$, respectively. The maximum reactive power setting $Q_{pv,k}^{max}$ is 60% of the apparent power rating of the PV unit $k$. The formulations of the Q-V characteristic (Fig. 3) are presented in (24)-(35). Note $z_{n,k} = 0$ indicates that the operating Zone $n$ is activated for PV unit $k$.

The if-then conditional statement for Zone 1 is given by:

$$u_k \leq v_1 + Mz_{1,k}, \forall k \in K \quad (24)$$

$$-Mz_{1,k} \leq Q_k^{pv} - Q_{pv,k}^{max} \leq Mz_{1,k}, \forall k \in K \quad (25)$$

The if-then conditional statement for Zone 2 is given by:

$$-Mz_{2,k} + v_1 \leq u_k \leq v_2 + Mz_{2,k}, \forall k \in K \quad (26)$$

$$-Mz_{2,k} \leq Q_k^{pv} - \frac{Q_{pv,k}^{max}}{v_2 - v_1}(v_2 - u_k) \leq Mz_{2,k}, \forall k \in K \quad (27)$$

The if-then conditional statement for Zone 3 is given by:

$$-Mz_{3,k} + v_2 \leq u_k \leq v_3 + Mz_{3,k}, \forall k \in K \quad (28)$$

$$-Mz_{3,k} \leq Q_k^{pv} \leq Mz_{3,k}, \forall k \in K \quad (29)$$

The if-then conditional statement for Zone 4 is given by:

$$-Mz_{4,k} + v_3 \leq u_k \leq v_4 + Mz_{4,k}, \forall k \in K \quad (30)$$

$$-Mz_{4,k} \leq Q_k^{pv} - \frac{Q_{pv,k}^{max}}{v_3 - v_4}(u_k - v_3) \leq Mz_{4,k}, \forall k \in K \quad (31)$$

The if-then conditional statement for Zone 5 is given by:

$$-Mz_{5,k} + \tilde{v}_4 \leq u_k, \quad \forall k \in K \quad (32)$$

$$-Mz_{5,k} \leq Q_k^{pv} + \tilde{Q}_{pv,k}^{max} \leq Mz_{5,k}, \quad \forall k \in K \quad (33)$$

The constraints (34)-(35) ensure that at least one operating zone is activated for each PV unit.

$$z_{1,k} + z_{2,k} + z_{3,k} + z_{4,k} + z_{5,k} \leq 4, \quad \forall k \in K \quad (34)$$

$$z_{1,k}, z_{2,k}, z_{3,k}, z_{4,k}, z_{5,k} \in \{0,1\}, \quad \forall k \in K \quad (35)$$

### C. Formulations of Q-V Characteristic of PV Units in ACOPF with Dispatch-VVC-Settings Co-optimization

The settings of the Q-V curve for PV units with VVC are considered as parameters based on default values of IEEE standard 1547-2018 in Section III.B. However, these settings can be adjusted for the VVC of each PV unit to meet the different needs of the system and improve the distribution grid operation. In this paper, the proposed ACOPF is extended to optimally identify the settings of VVC curves of smart inverters within the allowable range of IEEE standard 1547-2018. Formulations (36)-(45) show the additional constraint for the ACOPF with co-optimization of VVC curve settings, i.e., $\tilde{v}_{1,k}$, $\tilde{v}_{2,k}$, $\tilde{v}_{3,k}$, $\tilde{v}_{4,k}$, and $\tilde{Q}_{pv,k}^{max}$.

$$u_k \leq \tilde{v}_{1,k} + Mz_{1,k}, \quad \forall k \in K \quad (36)$$

$$-Mz_{1,k} \leq Q_k^{pv} - \tilde{Q}_{pv,k}^{max} \leq Mz_{1,k}, \quad \forall k \in K \quad (37)$$

$$-Mz_{2,k} + \tilde{v}_{1,k} \leq u_k \leq \tilde{v}_{2,k} + Mz_{2,k}, \quad \forall k \in K \quad (38)$$

$$-Mz_{2,k} \leq Q_k^{pv} - \frac{\tilde{Q}_{pv,k}^{max}}{\tilde{v}_{2,k} - \tilde{v}_{1,k}}(\tilde{v}_{2,k} - u_k) \leq Mz_{2,k}, \forall k \in K \quad (39)$$

$$-Mz_{3,k} + \tilde{v}_{2,k} \leq u_k \leq \tilde{v}_{3,k} + Mz_{3,k}, \quad \forall k \in K \quad (40)$$

$$-Mz_{3,k} \leq Q_k^{pv} \leq Mz_{3,k}, \quad \forall k \in K \quad (41)$$

$$-Mz_{4,k} + \tilde{v}_{3,k} \leq u_k \leq \tilde{v}_{4,k} + Mz_{4,k}, \quad \forall k \in K \quad (42)$$

$$-Mz_{4,k} \leq Q_k^{pv} - \frac{\tilde{Q}_{pv,k}^{max}}{\tilde{v}_{3,k} - \tilde{v}_{4,k}}(u_k - \tilde{v}_{3,k}) \leq Mz_{4,k}, \forall k \in K \quad (43)$$

$$-Mz_{5,k} + \tilde{v}_{4,k} \leq u_k, \quad \forall k \in K \quad (44)$$

$$-Mz_{5,k} \leq Q_k^{pv} + \tilde{Q}_{pv,k}^{max} \leq Mz_{5,k}, \quad \forall k \in K \quad (45)$$

The formulations (36)-(37), (38)-(39), (40)-(41), (42)-(43), and (44)-(45) are if-then conditional statements of co-optimization of the VVC curve settings for Zone 1, Zone 2, Zone 3, Zone 4, and Zone 5, respectively. However, making the settings of VVC as variables in the proposed model introduces non-convexity in the model due to two nonlinear terms $\frac{\tilde{Q}_{pv,k}^{max}}{\tilde{v}_{2,k}-\tilde{v}_{1,k}}(\tilde{v}_{2,k} - u_k)$ and $\frac{\tilde{Q}_{pv,k}^{max}}{\tilde{v}_{3,k}-\tilde{v}_{4,k}}(u_k - \tilde{v}_{3,k})$ from formulations (39) and (43). To handle the non-convexity issue in the model, the first-order Taylor series approximation is employed to convexify the nonlinear terms $\frac{\tilde{Q}_{pv,k}^{max}}{\tilde{v}_{2,k}-\tilde{v}_{1,k}}(\tilde{v}_{2,k} - u_k)$ and $\frac{\tilde{Q}_{pv,k}^{max}}{\tilde{v}_{3,k}-\tilde{v}_{4,k}}(u_k - \tilde{v}_{3,k})$ as two linearized formulations denoted as functions $f_2$ and $f_4$ shown in (46)-(47), respectively.

$$f_2(\tilde{Q}_{pv,k}^{max}, u_k, \tilde{v}_{1,k}, \tilde{v}_{2,k}) = \frac{v_{2,k}^{(t)} - u_k^{(t)}}{v_{2,k}^{(t)} - v_{1,k}^{(t)}} \tilde{Q}_{pv,k}^{max} - \frac{\tilde{Q}_{pv,k}^{max,(t)}}{v_{2,k}^{(t)} - v_{1,k}^{(t)}} u_k +$$
$$\frac{\tilde{Q}_{pv,k}^{max,(t)}(v_{2,k}^{(t)} - u_k^{(t)})}{(v_{2,k}^{(t)} - v_{1,k}^{(t)})^2} \tilde{v}_{1,k} - \frac{\tilde{Q}_{pv,k}^{max,(t)}(v_{1,k}^{(t)} - u_k^{(t)})}{(v_{2,k}^{(t)} - v_{1,k}^{(t)})^2} \tilde{v}_{2,k} \quad (46)$$

$$f_4(\tilde{Q}_{pv,k}^{max}, u_k, \tilde{v}_{3,k}, \tilde{v}_{4,k}) = \frac{u_k^{(t)} - v_{3,k}^{(t)}}{v_{3,k}^{(t)} - v_{4,k}^{(t)}} \tilde{Q}_{pv,k}^{max} - \frac{\tilde{Q}_{pv,k}^{max,(t)}}{v_{3,k}^{(t)} - v_{4,k}^{(t)}} u_k +$$
$$\frac{\tilde{Q}_{pv,k}^{max,(t)}(u_k^{(t)} - v_{4,k}^{(t)})}{(v_{3,k}^{(t)} - v_{4,k}^{(t)})^2} \tilde{v}_{3,k} - \frac{\tilde{Q}_{pv,k}^{max,(t)}(u_k^{(t)} - v_{3,k}^{(t)})}{(v_{3,k}^{(t)} - v_{4,k}^{(t)})^2} \tilde{v}_{4,k} \quad (47)$$

By substituting functions $f_2$ and $f_4$ into constraints (39) and (43), the linearized formulations of Zone 2 and Zone 4 for the ACOPF with co-optimization of VVC curve settings become formulations (48)-(49). In this paper, Taylor series base points in formulations (48)-(49) and (19)-(20) are updated at the same time. Moreover, the linearization errors of $f_2$ and $f_4$ and SOC relaxation errors are checked simultaneously in the proposed approach (Fig. 2).

$$-Mz_{2,k} \leq Q_k^{pv} - f_2(\tilde{Q}_{pv,k}^{max}, u_k, \tilde{v}_{2,k}, \tilde{v}_{3,k}) \leq Mz_{2,k}, \forall k \in K \quad (48)$$

$$-Mz_{4,k} \leq Q_k^{pv} - f_4(\tilde{Q}_{pv,k}^{max}, u_k, \tilde{v}_{3,k}, \tilde{v}_{4,k}) \leq Mz_{4,k}, \forall k \in K \quad (49)$$

Moreover, IEEE standard 1547-2018 provides allowable ranges for settings of VVC. The settings limits are considered and modeled in constraints (50)-(54).

$$0.82^2 \leq \tilde{v}_{1,k} \leq \tilde{v}_{2,k} - 0.04, \quad \forall k \in K \quad (50)$$

$$0.97^2 \leq \tilde{v}_{2,k} \leq 1, \quad \forall k \in K \quad (51)$$

$$1 \leq \tilde{v}_{3,k} \leq 1.03^2, \quad \forall k \in K \quad (52)$$

$$\tilde{v}_{3,k} + 0.04 \leq \tilde{v}_{4,k} \leq 1.18^2, \quad \forall k \in K \quad (53)$$

$$0 \leq \tilde{Q}_{pv,k}^{max} \leq S_{pv,k}^{max}, \quad \forall k \in K \quad (54)$$

With the proposed default settings of VVC (22)-(35) and optimal settings of VVC (22)-(23), (34)-(38), (40)-(42), (44)-(54), the proposed SOCP-based ACOPF model is extended into two MISOCP-based ACOPF models, which are tested on the real 1747-node unbalanced distribution system.

## IV. CASE STUDIES AND NUMERICAL RESULTS

### A. Test System Data and Assumptions

The proposed models are tested on an actual 1747-node three-phase unbalanced distribution network of a local electric utility in Arizona [23] with different feeders and laterals configurations, i.e., single-phase and three-phase. The total number of branches is 1744. The modeled load and PV generation data represent a snapshot at 2:00 pm on March 15, 2019. The instantaneous penetration of PV for this snapshot is 232%. The total active and reactive power demands are 1563.3 kW and 258.9 kVAr, respectively. The total active power injection from PV units is 3625.2 kW. An overview of this distribution feeder is shown in Fig. 4. In the original 1747-node distribution primary network shown, there is no PV unit with VVC installed. As a result, all PV units inject their maximum active power with unity power factor, i.e., zero reactive power, into the grid. This high penetration of PV units without VVC results in overvoltage issues, i.e., voltage magnitude exceeds 1.05 p.u., for some nodes at the studied snapshot. Furthermore, the testing system is modified by including 8 to 14 VVCs for the PV units at the nodes with the worst overvoltage issues. Two MISOCP-based ACOPF models, i.e., default and optimal settings of VVC, are tested on the modified distribution network.

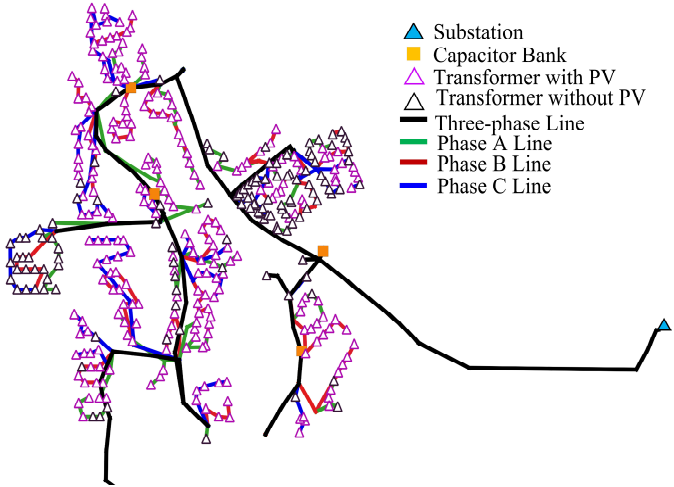

Fig. 4. Topology of 1747-node utility primary feeder.

The convex SOCP-based and MISOCP-based ACOPF models for the three-phase unbalanced system are implemented in Python and solved utilizing the Gurobi solver. The dynamic simulation is conducted using OpenDSS, a software tool developed by the Electric Power Research Institute (EPRI) [24] for distribution system studies. Simulations are conducted via a laptop with an Intel Core i7-10750H CPU, 16GB DDR4, and 1TB PCIe SSD. The PV generation price is modeled based on net surplus compensation rates of Pacific Gas and Electric (PG&E) on March 15, 2019 [25]. The wholesale electricity price is obtained from the locational marginal price (LMP) map of the California Independent System Operator (CAISO) for the PG&E area for March 15, 2019 [26]. In this paper, the PV generation and wholesale electricity prices are 21.79 \$/MWh and 55.71 \$/MWh, respectively, for the studied snapshot.

### B. Three-phase Unbalanced Convex SOCP-based ACOPF

The proposed SOCP-based ACOPF is tested on the system of Fig. 4 and the accuracy of the proposed model is evaluated in this section. Due to the excessive PV injection and overvoltage issue in the primary distribution feeder at the studied snapshot, the upper voltage limit of nodes is relaxed only in this section. Table I presents the maximum relaxation and linearization errors by applying the proposed convex SOCP-based ACOPF. Since the ACOPF convexifies formulation (9) into a SOC form in constraint (10), the SOC relaxation error $u_{i\phi}u_{j\phi'} - \left(c_{ij}^{\phi\phi'}\right)^2 - \left(s_{ij}^{\phi\phi'}\right)^2$ becomes a critical factor in checking the exactness of the obtained solution. The total simulation time for this test is 9.2 seconds.

For the 1747-node unbalanced network, the maximum SOC relaxation error is $1.1 \times 10^{-10}$ p.u., which is sufficiently small to ensure that the obtained solution is exact. The linearization errors of $c_{ij}^{\phi\phi'}$ and $s_{ij}^{\phi\phi'}$ need to be checked as well because Taylor series approximation is used to linearize the non-convex auxiliary variable formulations (19)-(20). Maximum linearization errors of $c_{ij}^{\phi\phi'}$ and $s_{ij}^{\phi\phi'}$ shown in Table I are considerably low, which implies that the approximations of two auxiliary variables are accurate enough in the proposed model to ensure an acceptable solution.

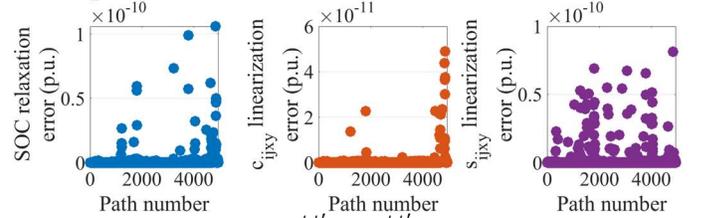

Fig. 5. SOC relaxation errors, $c_{ij}^{\phi\phi'}$ and $s_{ij}^{\phi\phi'}$ linearization errors of convex SOCP-based ACOPF for the 1747-node utility primary feeder.

Figure 5 presents the SOC relaxation errors, $c_{ij}^{\phi\phi'}$ and $s_{ij}^{\phi\phi'}$ linearization errors for all paths from node $i\phi$ (bus $i$ phase $\phi$) to node $j\phi'$ (bus $j$ phase $\phi'$) for the 1747-node three-phase unbalanced distribution primary feeder. In Fig. 5, most of the SOC relaxation errors are zero. A few of them are around $10^{-10}$ p.u.. The Taylor linearization errors of two auxiliary variables $c_{ij}^{\phi\phi'}$ and $s_{ij}^{\phi\phi'}$ are also small enough for all paths. Figure 5 indicates that the proposed convex SOCP-based ACOPF model can accurately handle relaxation and linearization errors and capture the three-phase unbalanced characteristics of a distribution system and obtain an exact global optimal solution.

### C. Results of MISOCP-based ACOPF with Q-V Characteristic of PV Units: Default and Optimal Settings

In this section, two MISOCP-based ACOPF models are tested using the modified 1747-node system with 8 VVCs for the nodes of PV units having the worst overvoltage issues. The location of 8 PV units with VVC for the modified 1747-node distribution primary feeder is shown in Fig. 6. The overvoltage issue is eliminated after adding VVCs because of the capability of providing reactive power support and curtailing active power from PV units. To further compare the performance of the default and optimal settings of VVC, two MISOCP-based ACOPF models are tested on the modified system with different number of VVCs.

TABLE I
MAXIMUM RELAXATION AND LINEARIZATION ERROR OF CONVEX SOCP-BASED ACOPF FOR THREE-PHASE UNBALANCED SYSTEM

| Case | Max SOC error (p.u.) | Max absolute linearization error for $c_{ij}^{\phi\phi'}$ (p.u.) | Max absolute linearization error for $s_{ij}^{\phi\phi'}$ (p.u.) |
|---|---|---|---|
| No VVC | $1.1 \times 10^{-10}$ | $4.9 \times 10^{-11}$ | $8.1 \times 10^{-11}$ |

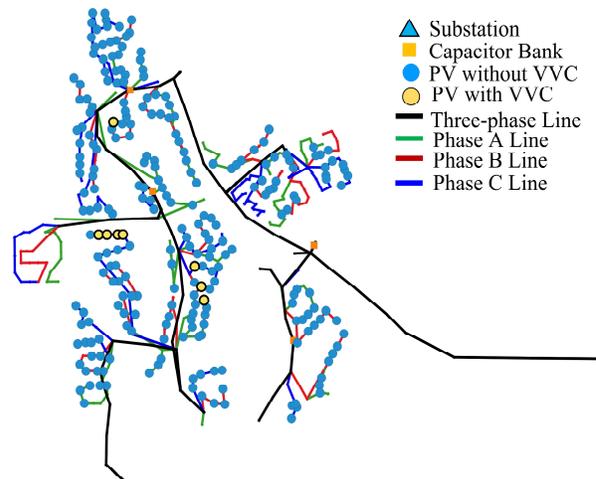

Fig. 6. Topology of 1747-node utility primary feeder with 8 VVC locations.

Since the proposed MISOCP-based ACOPF model also involves SOCP relaxation and Taylor series approximation, the relaxation errors and linearization differences need to be checked to ensure the exactness of the solution. The following results show the solutions of MISOCP-based ACOPF model for default and optimal settings of VVC.

TABLE II
MAXIMUM RELAXATION ERROR OF SOCP-BASED ACOPF WITH 8 PV UNITS WITH VVC

|  | Default settings | Optimal settings |
|---|---|---|
| Max SOC error (p.u.) | $1.14 \times 10^{-7}$ | $1.09 \times 10^{-7}$ |

TABLE III
MAXIMUM LINEARIZATION ERROR OF SOCP-BASED ACOPF WITH DEFAULT SETTINGS OF 8 VVCS

| Model | Max absolute linearization error for $c_{ij}^{\phi\phi'}$ (p.u.) | Max absolute linearization error for $s_{ij}^{\phi\phi'}$ (p.u.) |
|---|---|---|
| Default settings | $5.2 \times 10^{-8}$ | $9.8 \times 10^{-8}$ |

TABLE IV
MAXIMUM LINEARIZATION ERROR OF SOCP-BASED ACOPF WITH OPTIMAL SETTINGS OF 8 VVCS

| Model | Max absolute linearization error for $c_{ij}^{\phi\phi'}$ (p.u.) | Max absolute linearization error for $s_{ij}^{\phi\phi'}$ (p.u.) | Max linearization error for Q-V characteristic (p.u.) |
|---|---|---|---|
| Optimal settings | $5.0 \times 10^{-8}$ | $9.5 \times 10^{-8}$ | $1.2 \times 10^{-2}$ |

Table II shows the maximum SOCP relaxation errors of the proposed MISOCP-based ACOPF model for default and optimal settings of 8 PV units with VVCs. As shown in Table II, relaxation errors are small enough, which implies that the obtained solutions are exact for two MISOCP-based ACOPF models. The linearization errors of default and optimal settings models are shown in Tables III and IV, respectively. The linearization errors of $c_{ij}^{\phi\phi'}$ and $s_{ij}^{\phi\phi'}$ are all around $10^{-8}$ p.u.. It should be noted that due to high PV penetration and overvoltage issue in the studied snapshot, all 8 PV units with VVC operate in Zone 4 and absorb reactive power to mitigate the overvoltage issue. Therefore, only the linearization error of $f_4$ in the optimal settings model is shown in Table IV. It can be seen in Table IV that the maximum linearization error is $1.2 \times 10^{-2}$ p.u., which is small enough for a PV unit. Table III and Table IV imply that the proposed constraints based on Taylor approximations are accurate and sufficient to ensure acceptable solutions for both MISOCP-based ACOPF models.

Figures 7-8 present relaxation and linearization errors for all paths of the distribution network by applying the default and optimal settings models. The majority of SOC relaxation errors and $c_{ij}^{\phi\phi'}$ linearization errors are zero for both models. The $s_{ij}^{\phi\phi'}$ linearization errors are scattered from the maximum (aro-

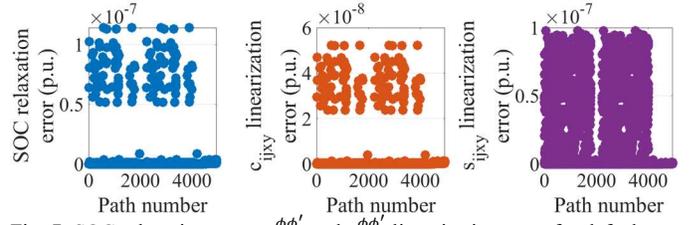
Fig. 7. SOC relaxation error, $c_{ij}^{\phi\phi'}$ and $s_{ij}^{\phi\phi'}$ linearization error for default settings model.

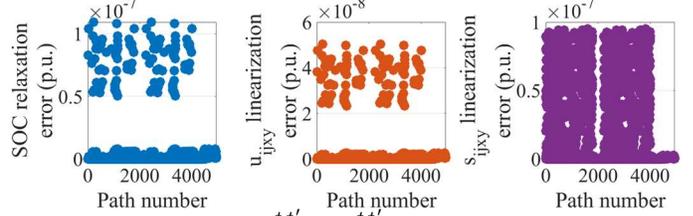
Fig. 8. SOC relaxation error, $c_{ij}^{\phi\phi'}$ and $s_{ij}^{\phi\phi'}$ linearization error for optimal settings model.

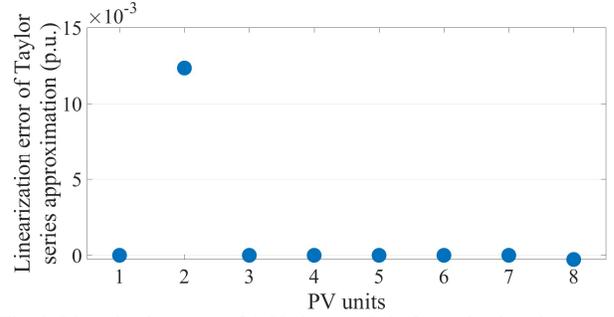
Fig. 9. Linearization error of Q-V characteristic for optimal settings model.

und $10^{-7}$) to zero. Figure 9 shows the linearization errors of $f_4$ for 8 PV units with VVCs in the ACOPF model with controller setting co-optimization. As it can be seen, only one PV unit has non-zero linearization error, which is relatively small. Figures 7-9 confirm that the proposed MISOCP-based ACOPF models for default and optimal settings of VVC can accurately represent the Q-V characteristic of VVC in the model and capture the characteristics of the unbalanced system.

Figures 10-11 illustrate the active and reactive power comparison of 8 PV units with VVCs for default and optimal setting models. Note that only 8 nodes of PV units having the worst overvoltage issues are selected to have VVC to mitigate the problems in the system. The system-level scheduling ensures that the PV units with VVC manage voltage at their local nodes and alleviate voltage violations across the feeder. Both models need to select a solution with active power curtailment and reactive power absorption from PV units with VVC to avoid overvoltage in the system due to the excessive PV active power injection. For the default settings model in Fig. 10, active power outputs of PV units 5, 6, and 8 are almost completely curtailed. However, there can be much less curtailment if settings are co-optimized in the model as shown in Fig. 10. Under default settings, there are active power curtailments for PV units with VVC in Fig. 10-11, while reactive power absorptions do not reach their maximum, i.e., $\tilde{Q}_{pv,k}^{max}$. This issue is due to the fixed characteristic of VVC for the default setting case. Zone 5 is activated if the node voltage reaches VVC voltage setting $V_4$, however, due to the voltage limit (i.e., 1.05 p.u.) being smaller

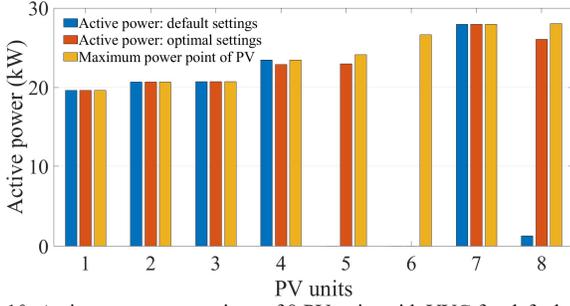
Fig. 10. Active power comparison of 8 PV units with VVC for default and optimal settings.

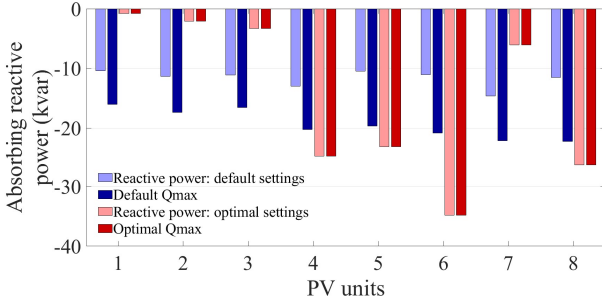
Fig. 11. Reactive power comparison of 8 PV units with VVC for default and optimal settings.

than the fixed VVC voltage setting $V_4$ (i.e., 1.06 p.u.), Zone 5 will not be activated under default settings. Thus, these VVCs operate in Zone 4 and the reactive power absorption is smaller than maximum reactive power absorption due to the fixed Q-V characteristic and settings (Fig. 3). In contrast, for the optimal settings model, the settings of VVC are adjusted to provide needed support to the distribution feeder, which results in less need for PV active power curtailment. The optimal settings of VVC enables the ACOPF model to attain more flexibility to adjust the reactive power output by changing the settings of VVC and facilitates reactive power support where it is most needed. Figures 10-11 confirms that co-optimizing the settings of VVC can result in more effective utilization of reactive power support by these units. The PV units 1, 2, 3, and 7 have less reactive power absorption, however, the rest of the PV units absorb more reactive power, which leads to less active power curtailments for the optimal settings model. The enhanced flexibility from optimal settings of VVCs results in an improvement of the system operation as well as a reduction in operational cost and PV active power curtailment as shown in Table V. The cost savings and percentage of the active power curtailments in this paper are calculated via (55) and (56), respectively.

$$\frac{\text{Cost difference between cases w/ default and optimal settings}}{\text{default settings cost}} \times 100\% \quad (55)$$

$$\frac{\text{Total active power curtailed from PV units with VVC}}{\text{Total active power of PV units with VVC}} \times 100\% \quad (56)$$

Case studies with different number of VVCs are conducted to compare the performance of default and optimal settings of VVC. To this end, more VVCs are considered for PV units of the nodes with the worst overvoltage issues. Table V shows the comparison and benefit of co-optimizing the settings of VVC within the ACOPF model. Note that the computational time is around 30 seconds for all simulations, and all obtained solutions are checked to have sufficiently small relaxation and linearization errors. As the number of VVCs increases, the PV curtailment decreases for both default and optimal settings models because more PV units can have higher ability to provide reactive power support. As shown in Table V, the default settings model always receives higher PV curtailment than the model with optimal settings of VVC. For instance, the optimal settings model reaches 0% PV curtailment, while the default settings model still has 16% PV curtailment in the case with 11 PV units with VVC. Also, the amounts of cost saving between the default and optimal settings models of VVC are shown in Table V. Due to the lower PV generation price compared to wholesale electricity price, less PV curtailment leads to less energy purchase from the more expensive wholesale electricity market. This implies that the system operating cost can be reduced by co-optimizing the settings of VVC as shown in Table V. Furthermore, the 9 VVCs case gains the highest cost-saving 7.4% by co-optimizing the settings in the ACOPF model.

TABLE V
COMPARISON OF DEFAULT AND OPTIMAL SETTINGS MODELS

| NO. of VVCs | Default settings | | Optimal settings | | Cost saving (%) |
|---|---|---|---|---|---|
| | Active power from PV (kW) | % Active power curtails | Active power from PV (kW) | % Active power curtails | |
| 8 | 113.8 | 41 | 158.1 | 17 | 6.8 |
| 9 | 153.9 | 28 | 196.5 | 8 | 7.4 |
| 10 | 182.8 | 23 | 229.1 | 3 | 6.3 |
| 11 | 217.9 | 16 | 259.0 | 0 | 6.1 |
| 12 | 238.5 | 13 | 274.1 | 0 | 5.7 |
| 13 | 268.4 | 10 | 294.2 | 0 | 5.4 |
| 14 | 306.4 | 6 | 326.9 | 0 | 4.0 |

### D. Dynamic Simulation for Stability Analysis of PV Unit: Optimal Settings

Since the improper selection of the VVCs' settings may cause system instability [1], dynamic analysis needs to be conducted to verify the control stability of PV units with VVC for optimal settings. In this paper, the stability of the optimal settings and the impact of load change disturbance are evaluated through dynamic analysis. The dynamic analyses are implemented via OpenDSS. The Q-V characteristic of PV units with VVC is programmed in a DLL file and called by OpenDSS. The result of the 8 VVCs case is exhibited as an exemplary dynamic simulation in this section. The time-domain of dynamic simulation is 0.9 sec. At the first 0.2 sec, the 8 VVCs of PV units are

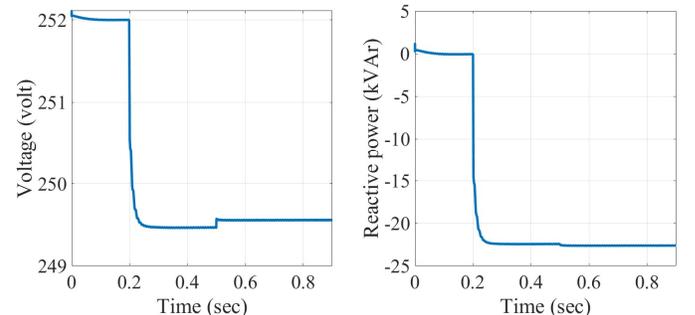
Fig. 12. Voltage and reactive power outputs of a PV unit with VVC.

inactivated. Then, all 8 VVCs with optimal settings are activated at 0.2 sec. Finally, load variation is modeled at 0.5 seconds of the time-domain simulation using Gaussian distribution with 20% load variance. The dynamic simulation results show that all 8 PV units with VVCs follow a similar pattern. Figure 12 presents the voltage and reactive power outputs of one PV unit with VVC.

As shown in Fig. 12, the voltage decreases, and reactive power output increases at 0.2 sec because of the activation of VVCs. Then, the load change has a very small impact on both curves. Therefore, both voltage and reactive power output curves become stable after a short period of time. This implies that the obtained optimal settings solution is acceptable and causes no stability issue for the system. Therefore, the response of the VVC is stable with the optimized settings.

## V. Conclusions

In this paper, a convex SOCP-based ACOPF model is proposed for three-phase unbalanced distribution systems. A two-stage iterative-based algorithm is developed to solve the proposed SOCP-based ACOPF. The Taylor series approximation is employed to create a linear relationship among the auxiliary variables to make the SOCP-based approach suitable for unbalanced distribution systems. The Q-V characteristics of PV units with VVC is considered based on the guidance of IEEE Standard 1547-2018. The proposed SOCP-based ACOPF is converted and extended into two MISOCP-based ACOPF models to account for the Q-V characteristic of PV units with VVC with default and optimal settings. All proposed models are tested on an actual 1747-node unbalanced distribution primary feeder.

To the best of the authors' knowledge, a convex SOCP-based ACOPF for the unbalanced distribution network is rarely discussed in the literature. This paper fills this research gap by proposing a convex SOCP-based ACOPF model for the three-phase unbalanced distribution grid operation while accounting for IEEE Standard 1547-2018 requirements for smart inverters. The proposed models can capture the characteristics of three-phase unbalanced network. The simulation results show that the proposed models can obtain the global optimal solution with very small relaxation and linearization errors, which means that the obtained solution is exact for the system. Moreover, the Q-V characteristic of PV units with VVC is well represented by the proposed MISOCP-based ACOPF. Besides, the results of the MISOP-based ACOPF show that the system operation can be improved if settings of VVC are co-optimized in the model due to the flexibility of the optimal settings model to adjust the reactive power output of PV units. Furthermore, dynamic analysis is conducted to ensure the stability of the system. The dynamic simulation results confirm that the optimal settings are valid and cause no stability concern for the distribution system.